\documentclass[letterpaper, 10 pt, conference]{ieeeconf}  

\IEEEoverridecommandlockouts                              
\usepackage{graphicx} 
\usepackage{amsmath} 
\usepackage{amssymb}  
\usepackage{caption}
\usepackage{subcaption}
\usepackage{balance}
\newtheorem{prop}{Proposition}

\title{\LARGE \bf
Multi-layer barrier function-based adaptive super-twisting controller}

\author{Antoine Thibault Vi{\'e}, Leonid Fridman, Roberto Galeazzi and Dimitrios Papageorgiou
\thanks{A. T. Vi{\'e}, R. Galeazzi and D. Papageorgiou are with the Department of Electrical and Photonics Engineering, Technical University of Denmark, Elektrovej 326, Kgs Lyngby, 2800, Denmark
        {\tt\small anthv,roga,dimpa@dtu.dk}}%
\thanks{L. Fridman is with Facultad de Ingenieria, Universidad Nacional Autonoma de Mexico,  Mexico City, 04510, Mexico
        {\tt\small lfridman@unam.mx}}%
}

\begin{document}

\maketitle
\thispagestyle{empty}
\pagestyle{empty}

\begin{abstract}
	This article presents an adaptive Super-Twisting Sliding Mode Control framework for uncertain  first-order systems, with rate-bounded perturbations, where the bound is constant but unknown. Positive definite barrier functions, when used in self-tuning super-twisting controllers may introduce some conservatism in relation to initial estimations of the perturbation rate bound. Moreover, discrete time implementation of the algorithm does not necessarily guarantee the boundedness of the closed-loop trajectories when sudden changes in the perturbation occur in between two time samples. The salient features of the proposed methodology pertain to extending the use of positive \emph{semidefinite} barrier functions to \emph{Super-Twisting} controller adaptation and the employment of a ``nested barriers" scheme that ensures boundedness of the solutions even for ``unfavourable" perturbations-to-sampling time ratios. The stability of the closed-loop system is assessed via Lyapunov analysis and simulations demonstrate the efficacy of the proposed framework.
\end{abstract}

\section{INTRODUCTION}
	Over the past decades Sliding Mode Control (SMC) methods have stood out in control systems design due to their robustness against uncertain perturbations. Conventional SMC algorithms rely on fixed parameters, which may not adapt well to varying conditions or objectives, making tuning difficult and limiting real-world applicability \cite{eisenzopf_adaptive_2021,papageorgiou_behaviour_2022}.

Self-tuning control strategies have been the object of recent studies to overcome these challenges. Adaptive methods adjust control parameters in real-time, enhancing system performance and resilience in changing environments \cite{plestan_new_2010,edwards_adaptive_2014}. Dynamic gain adaptation has been extended to higher-order SMC algorithms such as the Super-Twisting Sliding Mode Control (STSMC) \cite{shtessel_super-twisting_2010,moreno_adaptive_2015} based on a Lyapunov design or equivalent control approaches \cite{utkin_adaptive_2013}. One major limitation of such auto-tuning methods is the overestimation of the controller gains, especially when the bound on the perturbation or the perturbation derivative is unknown. Moreover, the ultimate bound on the closed-loop system state directly depends on the size of the perturbation compromising thus guarantees for predefined performance \cite{gonzalez_final_2024}.

Barrier function-based adaptation has recently emerged as a promising approach in addressing the conservativeness of conventional adaptive SMC methods \cite{obeid_barrier_2018,obeid2018adaptation} especially when the perturbation bound is not a priori known. Positive definite barrier functions (BF) were employed to ensure dynamic adaptation of the controller parameters ensuring convergence of the closed-loop trajectories within a predefined final set. The use of positive definite barrier functions was extended to STSMC \cite{obeid_barrier_2020} and higher-order SMC \cite{laghrouche_barrier_2021} to enhance control robustness and constraint enforcement. Positive semi-definite barrier functions, were explored in \cite{cruz2021barrier} for first-order SMC and in \cite{gonzalez_final_2024,ovalle2025analysis} for signed power-based and bi-powered control laws, contributing to reduced conservativeness of the controller gains near the boundary of the barrier set.

When implementing barrier function-based adaptive schemes, the regions outside the barrier function’s
domain must be carefully studied to maintain control performance across the entire operational range of the system. This becomes more apparent when translating continuous-time control strategies into discrete-time systems for digital implementation. It has been shown \cite{ovalle2025discrete} that predefined performance under sampling is not achievable irrespective of the controller type. This is due to the implicit assumptions of infinite actuator capacity and bandwidth, both unrealistic in actual systems.

While the authors in \cite{ovalle2025discrete} derived a relation between allowable perturbation and sampling rate of the discrete SMC algorithm and introduced a reformulation of the main control objective, this paper proposes an alternative approach to ensure predefined performance based on pre-definition of several "sub-optimal" accuracy levels and the use of multiple nested barrier functions within a STSMC framework. This multi-layer architecture can enhance control performance by providing distinct control objectives depending on the (possibly changing) perturbation bound
. The contributions of this work specifically pertain to:
\begin{itemize}
	\item The use of positive \emph{semidefinite} barrier functions for the adaptive non-homogeneous super-twisting scheme instead of a signed power-based and bi-powered control laws. This approach combines the feature of reduced conservativeness in the gain adaptation with the enhanced robustness of controllers such as the STSMC. The stability of the closed-loop system is analysed.
	
	\item A Multiple Barrier Functions scheme in a layered architecture that can handle different perturbation rates and provide tighter bounds on the trajectories without exclusively relying on gain adaptation, the latter leading to excessive actuation. This architecture is very essential in discrete-time implementations, when the effect of the perturbation on the system between two samples can force the trajectories out of the barrier set. 
\end{itemize}

The remainder of the paper is organised as follows: Section \ref{sec:problemformulation} describes the problem at hand. Section \ref{sec:BFAdapation} presents the BF-based adaptive non-homogeneous STSMC with positive semidefinite functions and Section \ref{sec:MultiLayer} details the multi-layered BF-scheme. Simulation results are presented in Section $\sec$ and finally, conclusions and elements of future work are discussed in Section \ref{sec:conclusions}.

\section{PROBLEM FORMULATION} \label{sec:problemformulation}
	The system under study is a perturbed integrator:
\begin{equation}
\label{eq:system}
	\dot{s} = u + d(t)
\end{equation}
where $s,u \in \mathbb{R}$ and $d : \mathbb{R}^+ \rightarrow \mathbb{R}$. Any system of relative degree $\varrho$ can be described by \eqref{eq:system} based on an appropriate selection of a sliding-variable (corresponding to an $(\varrho - 1)$-dimensional sliding manifold) and equivalent control \cite{slotine1991applied}.
According to the non-homogeneous STSMC and using the shorthand notation $\lfloor s \rceil^a \triangleq \vert s \vert^a\text{sgn}(s)$, the control input $u$ is expressed as:
\begin{subequations} \label{eq:nonHomogeneousSTSMC}
	\begin{align}
		u &= -k_1 \lfloor s \rceil^\alpha + v \\
		\dot{v} &= -k_2 \lfloor s \rceil^0  
	\end{align}
\end{subequations}
where $\alpha \in \mathbb{R}^+$. The algorithm is homogeneous only for $\alpha = \frac{1}{2}$, \cite{moreno_strict_2014}.
The closed-loop system, whose solutions are to be understood in the sense of Filippov, is written as:
\begin{subequations}
	\begin{align}
		\dot{s} &= -k_1 \lfloor s \rceil^\alpha + \phi \\
		\dot{\phi} &= -k_2 \lfloor s \rceil^0  + \delta(t)
	\end{align}
	\label{eq:STA}
\end{subequations}
\noindent where $\delta(t) \triangleq \dot{d}(t)$ is bounded, i.e. $\exists \Delta > 0 \text{ s.t } \forall t \in  \mathbb{R}^+, \lvert \delta(t) \rvert \leq \Delta$ and $\Delta$ is a priori unknown.
It has been demonstrated in \cite{moreno_strict_2014} that for $\alpha \in \left[0,\frac{1}{2}\right]$ and with appropriately chosen fixed $k_1$ and $k_2$  the system described by \eqref{eq:STA} can achieve semi-globally finite time stability. However, ensuring invariance of the origin to non-vanishing bounded perturbations for $\frac{1}{2} < \alpha < 1$ and $\alpha > 1$ has not been demonstrated for the STSMC algorithm. On the other hand, it has been shown \cite{gonzalez_final_2024} that for the conventional SMC, employing barrier functions ensures Uniform Ultimate Boundedness (UUB) of the system trajectories, for any $\alpha \in \mathbb{R}^+$.



\section{Barrier function-based adaptive STSMC} \label{sec:BFAdapation}
	This section presents a two-mode gain adjustment framework for the non-homogeneous STSMC in \eqref{eq:nonHomogeneousSTSMC}. The first mode is based on barrier functions and concerns operation within a pre-defined set. The second mode pertains to dynamic adaptation outside the barrier, applicable to large initial conditions and sudden perturbations in combination to slow sampling.

\subsection{Adaptation inside the barrier}
\begin{figure}[h!]
	\centering
	\includegraphics[width = 0.45\textwidth, height=4.5cm]{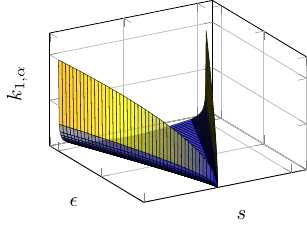}
	\caption{$k_{1,\alpha}(s,\epsilon)$ for $\epsilon \in \left[1e^{-4},1e^{-1}\right], \lvert s \rvert < 0.99\epsilon$, $\alpha = \frac{1}{2}$}
	\label{fig:k_s_epsilon}
\end{figure}

	Consider the following barrier functions \cite{gonzalez_final_2024}, \cite{obeid_barrier_2019}:
\begin{align} \label{eq:barrier_function}
	k_{1,\alpha}(s,\epsilon) &= \frac{\lvert s \rvert}{\left(\epsilon - \lvert s \rvert\right)^{\alpha + 1}}, \; k_{2,\alpha}(s,\epsilon) = k_{1,\alpha}(s,\epsilon)^2
\end{align}
with $\epsilon$ and $\alpha \in \mathbb{R}^+$. Hence, according to \cite{gonzalez_final_2024}, $k_{1,\alpha} : \left]-\epsilon,\epsilon \right[ \times \mathbb{R}^+ \rightarrow \mathbb{R}^+$ fulfill every requirement to be a barrier function, therefore $k_{2,\alpha} : \left]-\epsilon,\epsilon \right[ \times \mathbb{R}^+ \rightarrow \mathbb{R}^+$ is also a barrier function. Therefore, as depicted in Fig.\ref{fig:k_s_epsilon}, $k_{1,\alpha}$ behaves analogously to a potential well, ensuring that the sliding variable remains within the desired accuracy and cannot be pushed out.
\begin{prop}
	 The control law \eqref{eq:nonHomogeneousSTSMC} with $k_1 = k_{1,\alpha}(s,\epsilon)$, $k_2 = k_{2,\alpha}(s,\epsilon)$ ensures that the trajectories of the closed loop system \eqref{eq:STA}, once started in the set $\vert s \vert < \epsilon$, they will remain there for all future times.
\end{prop}
\begin{proof} Define the set
\begin{equation}
	\Pi \overset{\Delta}{=} \left[-\zeta, -\theta\right]\cup \left[\theta, \zeta\right] \subset \left] -\epsilon, \epsilon \right[
\end{equation}
and the variables $y_{1,\alpha} = k_{1,\alpha}^2 s$ and $y_{2,\alpha} = \phi$ for $s\in\Pi$. Note that the dependency of these functions on $s$ and $\epsilon$ is omitted for brevity. The dynamic system with these new variables is given by:
	\begin{align} \label{eq:Modified_STA}
		\dot{y}_{1,\alpha} &= 2 \frac{\dot{k}_{1,\alpha}}{k_{1,\alpha}} y_{1,\alpha} + k_{1,\alpha}^2 \dot{s} \text{ and } \dot{y}_{2,\alpha} = \dot{\phi}
	\end{align}
Noting that
\begin{subequations}
	\begin{align}
		\frac{\dot{k}_{1,\alpha}}{k_{1,\alpha}} &= h_{\alpha}(s,\epsilon)k_{1,\alpha}^2 \lfloor s \rceil^0 \dot{s} \\
		h_{\alpha}(s,\epsilon) &= \frac{\left(\epsilon + \alpha \lvert s \rvert \right) \left(\epsilon - \lvert s \rvert\right)^{2\alpha+1}}{\lvert s \rvert^3}
	\end{align}
\end{subequations} 
where $h_\alpha: \Pi \times \mathbb{R}^+ \rightarrow \mathbb{R}^+$, allows using arguments inspired by the proof in \cite{obeid_barrier_2019}.
Indeed, \eqref{eq:Modified_STA} can be rewritten as:
\begin{subequations}
	\begin{align}
		\frac{\dot{y}_{1,\alpha}}{1 + 2h_\alpha(s,\epsilon) \lvert y_{1,\alpha} \rvert} &= k_{1,\alpha}^2\left(-\lfloor y_{1,\alpha} \rceil^\alpha + y_{2,\alpha} \right) \\
		\dot{y}_{2,\alpha} &= -k_{1,\alpha}^2 \lfloor y_{1,\alpha} \rceil^0 + \delta(t)
	\end{align}
	\label{eq:STA_proof}
\end{subequations}
It is assumed that $\exists \bar{t}(s(0)) \text{ s.t. } \lvert s(\bar{t}) \rvert \leq \frac{\epsilon}{2}$. This assumption will be proven in the next section when $s(0) \geq \frac{\epsilon}{2}$. Subsequently, a new time scale $\tau$ is defined as $\tau(\bar{t}(s(0))  = 0$ and $d\tau = k_{1,\alpha}^2 dt$. This new time scale begins only after the previous assumption has been fulfilled. The symbol $\prime$ is used to denote the derivative concerning $\tau$. Hence, \eqref{eq:STA_proof} can be rewritten as:
\begin{subequations}
	\begin{align}
		\frac{y_{1,\alpha}^\prime}{1 + 2h_\alpha(s,\epsilon) \lvert y_{1,\alpha} \rvert} &=-\lfloor y_{1,\alpha} \rceil^\alpha + y_{2,\alpha} \\
		y_{2,\alpha}^\prime &= \lfloor y_{1,\alpha} \rceil^0 + \frac{\delta(t)}{k_{1,\alpha}^2}
	\end{align}
	\label{eq:STA_proof_tau}
\end{subequations}

Before defining the Lyapunov function for the stability analysis, it is important to note that $h_{\alpha}$ is even and monotonically decreasing on $\left[\theta,\zeta\right]$, hence two constants are defined:
\begin{equation}
	C_{\theta,\alpha} = h_\alpha(\theta,\epsilon) \geq h_\alpha(s,\epsilon) \geq h_\alpha(\zeta,\epsilon) = C_{\zeta,\alpha} \mbox{, } \forall s \in \Pi
\end{equation}

It is important to note that
\begin{align}
	\frac{\delta}{k_{1,\alpha}^2} &\leq \frac{\Delta}{k_{1,\alpha}^2} = \frac{\Delta \left(\epsilon - \lvert s \rvert\right)^{2\alpha+1}\left(3\epsilon+(2\alpha-1)\lvert s \rvert\right)}{\lvert s \rvert^2 \left( 1 + 2h_\alpha(s,\epsilon)\lvert y_{1,\alpha} \rvert \right)} \nonumber\\
	&= \Omega_\alpha(s,\epsilon) \frac{\Delta}{1+2h_\alpha(s,\epsilon) \lvert y_1 \rvert}
	\label{eq:DisturbanceInequality}
\end{align}
Subsequently, the following Lyapunov function is defined:
\begin{align}
	V &= \frac{\ln(1 + 2C_{\zeta,\alpha}\lvert y_{1,\alpha} \rvert)}{2 C_{\zeta,\alpha}}\left(1 - \frac{1}{4}\lfloor y_{1,\alpha} \rceil^0 \sigma(y_{2,\alpha})\right) \nonumber\\
	&+ \frac{F(y_{1,\alpha},y_{2,\alpha}) y_{2,\alpha}^2}{2}
\end{align}
where $\sigma(y_{2,\alpha})$ is a saturation function defined as
\begin{equation}
	\sigma(y_{2,\alpha}) = \lfloor y_{2,\alpha} \rceil^0 \min \left(\lvert y_{2,\alpha} \rvert,1\right)
	\label{eq:sigma}
\end{equation}
and $F(y_{1,\alpha},y_{2,\alpha}) = L_\alpha > 1$ if $\lfloor y_{1,\alpha} \rceil^0 y_{2,\alpha} \leq 0$ and $F(y_{1,\alpha},y_{2,\alpha}) = 1$ if $\lfloor y_{1,\alpha} \rceil^0 y_{2,\alpha} > 0$. The following inequality holds for every $\left(y_{1,\alpha},y_{2,\alpha}\right) \in \mathbb{R}^2$:
\begin{align}
	&\frac{3 \ln \left(1 + 2 C_{\zeta,\alpha} \lvert y_{1,\alpha} \rvert\right)}{8 C_{\zeta,\alpha}} + \frac{y_{2,\alpha}^2}{2} \leq V \nonumber\\
	&\leq \frac{5 \ln \left(1 + 2 C_{\zeta,\alpha} \lvert y_{1,\alpha} \rvert\right)}{8 C_{\zeta,\alpha}} + \frac{L_\alpha y_{2,\alpha}^2}{2}
\end{align}
Hence, $V$ is positive-definite, decrescent, and radially unbounded. Moreover of class $C^1$ if $y_{1,\alpha} \neq 0$ and $\lvert y_{2,\alpha} \rvert \geq 1$. Subsequently, according to \cite{obeid_barrier_2019}, \eqref{eq:sigma} and \eqref{eq:DisturbanceInequality}, one gets
\begin{align*}
	\left(\sigma(y_{2,\alpha})\right)^\prime &\leq \sigma(y_{2,\alpha})^\prime \frac{\Delta}{k_{1,\alpha}^2} = \sigma(y_{2,\alpha})^\prime \frac{\Omega_\alpha(s,\epsilon) \Delta}{1+2h_\alpha(s,\epsilon) \lvert y_1 \rvert}
\end{align*}
\begin{equation}
\begin{split}
	V^\prime &= -\overbrace{\frac{1 + 2 h_\alpha(s,\epsilon)\lvert y_{1,\alpha} \rvert}{1 + 2C_\zeta \lvert y_{1,\alpha} \rvert}}^{M_\alpha(s,\epsilon)}\lfloor y_{1,\alpha} \rceil^0 \lfloor y_{1,\alpha} \rceil^\alpha \\&+ \frac{1 + 2 h_\alpha(s,\epsilon)\lvert y_{1,\alpha} \rvert}{1 + 2C_\zeta \lvert y_{1,\alpha} \rvert}\lfloor y_{1,\alpha} \rceil^0 y_{2,\alpha} \\&- F(y_{1,\alpha},y_{2,\alpha})\lfloor y_{1,\alpha} \rceil^0 y_{2,\alpha} +  F(y_{1,\alpha},y_{2,\alpha}) \frac{\delta(t)}{k_{1,\alpha}^2}y_{2,\alpha} \\&- \frac{1 + 2 h_\alpha(s,\epsilon)\lvert y_{1,\alpha} \rvert}{4\left(1 + 2C_\zeta \lvert y_{1,\alpha} \rvert\right)} \sigma(y_{2,\alpha}) y_{2,\alpha} \\ &-\left(\sigma(y_{2,\alpha})\right)^\prime \lfloor y_{1,\alpha} \rceil^0 \frac{\ln \left(1 + 2 C_{\zeta,\alpha}\right)}{8 C_{\zeta,\alpha}}
\end{split}
\end{equation}
with $1 \leq M_\alpha(s,\epsilon) \leq L_{\alpha},  \forall s\in\Pi$. Therefore
\begin{align*}
		&V^\prime \leq -\frac{3 M_\alpha(s,\epsilon)}{4} \lvert y_{1,\alpha} \rvert^\alpha \nonumber\\
		&- \lvert y_{2,\alpha} \rvert \left(\frac{M_\alpha(s,\epsilon)}{4} - \Omega_\alpha(s,\epsilon) \frac{\Delta}{1 + 2C_{\zeta,\alpha} \lvert y_{1,\alpha} \rvert}\right) \nonumber\\ &+ \sigma(y_{2,\alpha})^\prime \Omega_\alpha(s,\epsilon) \frac{\Delta}{8 C_{\zeta,\alpha}}\frac{\ln \left(1 + 2C_{\zeta,\alpha} \lvert y_{1,\alpha} \rvert \right)}{1 + 2C_{\zeta,\alpha} \lvert y_{1,\alpha} \rvert} \nonumber\\
		&+ \left(M_\alpha(s,\epsilon) - F(y_{1,\alpha},F_{2,\alpha})\right)\lfloor y_{1,\alpha} \rceil^0 y_{2,\alpha}
\end{align*}
Note that a non-trivial trajectory of \eqref{eq:Modified_STA} crosses the line $y_{1,\alpha} = 0$ in isolated points. Moreover, at time $\tau$ where $y(\tau) = (0,y_{2,\alpha}(\tau))$ with $y_{2,\alpha}(\tau) > 0$, V admits an isolated discontinuity jumping from $y_{2,\alpha}^2$ to $L_\alpha y_{2,\alpha}^2$ and conversely if $y_{2,\alpha}(\tau) < 0$. Finally, note that outside the aforementioned line and for $\lvert y_{2,\alpha} \rvert \geq 1$, V is a $C^1$ function of the time $\tau$ verifying:
\begin{align*}
	V^\prime &\leq -\frac{3 M_\alpha(s,\epsilon)}{4} \lvert y_{1,\alpha} \rvert^\alpha \nonumber\\
	&- \lvert y_{2,\alpha} \rvert \left(\frac{M_\alpha(s,\epsilon)}{4} - \Omega_\alpha(s,\epsilon) \frac{\Delta}{1 + 2C_{\zeta,\alpha} \lvert y_{1,\alpha} \rvert}\right) \nonumber\\
	&+ \sigma(y_{2,\alpha})^\prime \Omega_\alpha(s,\epsilon) \frac{\Delta}{8 C_{\zeta,\alpha}}\frac{\ln \left(1 + 2C_{\zeta,\alpha} \lvert y_{1,\alpha} \rvert \right)}{1 + 2C_{\zeta,\alpha} \lvert y_{1,\alpha} \rvert}
\end{align*}
It is easily proven that $\lvert s \rvert$ remains lower than $\epsilon$, indeed:
\begin{align*}
	\lim_{\lvert s \rvert \rightarrow \zeta, \zeta\rightarrow\epsilon} V^\prime = \lim_{\lvert s \rvert \rightarrow \epsilon} V^\prime = \lim_{y_{1,\alpha} \rightarrow \epsilon} V^\prime = -\frac{3}{4}\lvert y_{1,\alpha} \rvert^\alpha - \frac{1}{4} \lvert y_{2,\alpha} \rvert
\end{align*}
i.e $\left(V^\frac{1}{2}\right)^\prime \leq -C$ for some positive constant. Then one would have convergence in finite time to 0, which is impossible.

Subsequently, it is demonstrated that the proof of Theorem 1 in \cite{obeid_barrier_2019} also holds for this barrier function. Indeed, since:
\begin{align*}
	&\lim_{\lvert s \rvert \rightarrow \epsilon} \Omega_\alpha(s,\epsilon) = 0 \text{ and } \lim_{\lvert s \rvert \rightarrow 0} \Omega_\alpha(s,\epsilon) = \infty
\end{align*}
one gets, $\forall b \in \mathbb{R^{*+}}, \exists s^* \in \Pi$ such as the following holds:
\begin{align*}
	\Omega_\alpha(s,\epsilon) &\leq \frac{1}{b^2} \mbox{  } \forall \lvert s \rvert \geq \lvert s^* \rvert
\end{align*}
The function $\Omega_\alpha$ can be interpreted as a variable gain that dynamically adjusts to manage the effect of the disturbance on the system. This adjustment is based on the value of the sliding variable. Specifically, $\Omega_\alpha$  modulates the system's response to disturbances within the predefined constraints of the control design. By doing so, it ensures that the control action is neither overly aggressive nor too passive, thus maintaining system stability and performance.
Hence, by defining $b^* = \frac{1}{\sqrt{\Omega_\alpha(s^*,\epsilon)}}$, $\forall \lvert s \rvert \geq s^*$ one gets:
\begin{align}
	V^\prime &\leq -\frac{3M_\alpha(s,\epsilon)}{4}\lvert y_1 \rvert^\alpha \nonumber\\
	&- \lvert y_2 \rvert \left(\frac{M_\alpha(s,\epsilon)}{4} - \frac{1}{(b^*)^2} \frac{\Delta}{1 + 2C_{\zeta,\alpha}\lvert y_{1,\alpha} \rvert}\right) \nonumber\\
	&+ \sigma(y_{2,\alpha})^\prime \frac{\Delta}{8(b^*)^2C_{\zeta,\alpha}} \frac{\ln \left(1 + 2C_{\zeta,\alpha} \lvert y_{1,\alpha} \rvert \right)}{1 + 2C_{\zeta,\alpha} \lvert y_{1,\alpha} \rvert }
\end{align}
Moreover, since $\min M_\alpha(s,\epsilon) = 1$:
\begin{align*}
	V^\prime &\leq -\frac{3}{4}\lvert y_1 \rvert^\alpha - \lvert y_2 \rvert \left(\frac{1}{4} - \frac{1}{(b^*)^2} \frac{\Delta}{1 + 2C_{\zeta,\alpha}\lvert y_{1,\alpha} \rvert}\right) \nonumber\\
	&+ \sigma(y_{2,\alpha})^\prime \frac{\Delta}{8(b^*)^2C_{\zeta,\alpha}} \frac{\ln \left(1 + 2C_{\zeta,\alpha} \lvert y_{1,\alpha} \rvert \right)}{1 + 2C_{\zeta,\alpha} \lvert y_{1,\alpha} \rvert } \; .
\end{align*}
Furthermore, since this holds for every $s^* \in \Pi$, accordingly to an appropriately chosen $b$, the same arguments from Theorem 1 of \cite{obeid_barrier_2019} can be used to conclude the boundedness of the trajectories within the barrier set.
\end{proof}

\subsection{Adaptation outside the barrier}
	The second mode of the adaptive framework pertains to the operation outside the barrier set. The main result is summarised in the subsequent proposition.
\begin{prop}
	The control law \eqref{eq:nonHomogeneousSTSMC} with $k_1 = k_{1,\alpha}(s,\epsilon)$, $k_2 = k_{2,\alpha}(s,\epsilon)$ and
	\begin{subequations}
		\begin{align}
			\dot{k}_{1,\alpha} &= \begin{cases}
				\frac{k_{1,\alpha}}{\lvert \dot{s} \rvert} \mbox{ if } \lvert s \rvert \geq \epsilon \\
				-k_{1,\alpha} \mbox{ otherwise}
			\end{cases} \\
			\dot{k}_{2,\alpha} &= \begin{cases}
				\frac{k_{2,\alpha}}{2 \lvert s \rvert^{1-\alpha}} \mbox{ if } \lvert s \rvert \geq \epsilon \\
				-k_{2,\alpha} \mbox{ otherwise}
			\end{cases}
		\end{align}
		\label{eq:STA_dynamics}
	\end{subequations}
	ensures that the trajectories of the closed loop system \eqref{eq:STA} converge to the set $\vert s \vert < \epsilon$ in finite time.
\end{prop}

\begin{proof}
Define the Lyapunov function \cite{moreno_strict_2014}
\begin{equation}
	V = z_\alpha^T R z_\alpha + 2 \gamma k_{2,\alpha} \lvert s \rvert
\end{equation}
with $\gamma \in \mathbb{R}^{*+}$ and:
\begin{align*}
	z_\alpha &= \begin{bmatrix}
		\lfloor s \rceil^\alpha & \psi
	\end{bmatrix}^T, \; \psi = k_{1,\alpha} \lfloor s \rceil^\alpha - \phi = -\dot{s}\\
	R &= \frac{\gamma}{2} \begin{bmatrix}
		k_{1,\alpha}^2 & -k_{1,\alpha} \\
		-k_{1,\alpha} & 2
	\end{bmatrix}
\end{align*}
Since, $\gamma$ and $k_{1,\alpha}$ are positive and $R$ is positive definite,
\begin{align*}
	\lambda_{min}(R) \{\lvert s \rvert^{2\alpha} + \psi^2\} &\leq V - 2 \gamma k_{2,\alpha} \lvert s \rvert \\
	&\leq \lambda_{max}(R) \{\lvert s \rvert^{2\alpha} + \psi^2\}
\end{align*}
showing that $V$ is positive definite, decrescent, and radially unbounded.
Moreover, $V$ is continuous and continuously differentiable almost everywhere. Hence, $\forall (s,\phi) \in \mathbb{R}^2 \wedge s \neq 0$, the time derivative of $V$, is given as
\begin{align}
	\dot{V} &= -\gamma k_{1,\alpha} k_{2,\alpha} \lvert s \rvert^\alpha - \frac{\gamma}{2} \alpha k_{1,\alpha} \psi^2 \lvert s \rvert ^{\alpha-1} + \gamma \dot{k}_{1,\alpha} \psi \lfloor s \rceil^\alpha \nonumber\\
	&+ 2 \gamma \dot{k}_{2,\alpha} \lvert s \rvert + \gamma k_{1,\alpha} \lfloor s \rceil^\alpha \delta(t) - 2 \gamma \psi \delta(t)
\end{align}
consequently, replacing the gain derivative by their expressions given in \eqref{eq:STA_dynamics}:
\begin{align}
	\dot{V} &= -\gamma k_{1,\alpha} k_{2,\alpha}\lvert s \rvert^\alpha - \frac{\gamma}{2} \alpha  k_{1,\alpha} \psi^2 \lvert s \rvert ^{\alpha-1} + \gamma \frac{k_{1,\alpha}}{\lvert \psi \rvert} \psi \lfloor s \rceil^\alpha \nonumber\\
	&+ 2\gamma \frac{k_{2,\alpha}}{2 \lvert s \rvert^{1-\alpha}} \lvert s \rvert + \gamma k_{1,\alpha} \lfloor s \rceil^\alpha \delta(t) - 2 \gamma \psi \delta(t)
\end{align}
If $\alpha \in \left]0,1\right[$ the Beckenback-Bellman Lemma \cite{beckenbach2012inequalities} can be used and then, $\forall \mu \in \mathbb{R}^{*+}$
\begin{equation}
	\lvert s \rvert^{\alpha-1} \psi^2 \geq -\frac{\left(1-\alpha\right)\mu^\frac{-\alpha}{1-\alpha}}{\alpha}\lvert s \rvert^\alpha + \frac{\mu^{-\alpha}}{\alpha}\lvert \psi \rvert^{2\alpha}
\end{equation}
Therefore:
\begin{align}
	\dot{V} &\leq -\gamma \alpha k_{1,\alpha} \left(\frac{\mu^{-\alpha}}{\alpha}\lvert \psi \rvert^{2\alpha} -\frac{\left(1-\alpha\right)\mu^\frac{-\alpha}{1-\alpha}}{\alpha}\lvert s \rvert^\alpha \right) \nonumber\\
	&+ \gamma k_{2,\alpha} \lvert s \rvert^\alpha - \gamma k_{1,\alpha}k_{2,\alpha} \lvert s \rvert^\alpha + \gamma k_{1,\alpha} \lvert s \rvert^\alpha \Delta \nonumber\\
	&+ 2 \gamma \lvert \psi \rvert \Delta + \gamma k_{1,\alpha}\lvert s \rvert^\alpha
	\label{eq:Lyap_Dynamics}
\end{align}
\underline{In the unperturbed case ($\Delta = 0$)}, \eqref{eq:Lyap_Dynamics} can be written as:
\begin{align}
	\dot{V} &\leq -\gamma \alpha k_{1,\alpha} \left(-\frac{\left(1-\alpha\right)\mu^\frac{-\alpha}{1-\alpha}}{\alpha}\lvert s \rvert^\alpha + \frac{\mu^{-\alpha}}{\alpha}\lvert \psi \rvert^{2\alpha}\right) \nonumber\\
	&- k_{1,\alpha}k_{2,\alpha} \lvert s \rvert^\alpha + \gamma k_{1,\alpha}\lvert s \rvert^\alpha + \gamma k_{2,\alpha} \lvert s \rvert^\alpha
\end{align}
And therefore:
\begin{align}
	\dot{V} &\leq -\gamma \left(k_{1,\alpha} \left(k_{2,\alpha} - \left(1-\alpha\right)\mu^\frac{-\alpha}{1-\alpha} - 1\right) - k_{2,\alpha} \right)\lvert s \rvert^\alpha \nonumber\\
	&- \gamma k_{1,\alpha} \mu^{-\alpha}\lvert \psi \rvert^{2\alpha}
\end{align}
By selecting $\mu$ as:
\begin{equation}
	\mu = \eta \left(\frac{(1- \alpha)k_{1,\alpha}}{k_{1,\alpha}\left(k_{2,\alpha }-1\right) - k_{2,\alpha}}\right)^{\frac{1-\alpha}{\alpha}} \text{ } \eta > 1
\end{equation}
the first coefficient in the last expression is positive everywhere, except at the singularity $k_{1,\alpha} = \frac{k_{2,\alpha}}{k_{2,\alpha} - 1}$ where $\mu$ not defined. In order to avoid this specific case, if $k_{1,\alpha} = \frac{k_{2,\alpha}}{k_{2,\alpha} - 1}$ then a value $\nu > 0$ is added to $k_{1,\alpha}$. Hence, $\dot{V}$ is rewritten as:
\begin{equation}
	\dot{V} \leq -\rho_1 \lvert s \rvert^\alpha -\rho_2 \lvert \psi \rvert^{2\alpha}
\end{equation}
with
\begin{equation}
	\begin{split}
		\rho_1 &= \gamma \left( k_{1,\alpha} \left(k_{2,\alpha} - \left(1-\alpha\right)\mu^\frac{-\alpha}{1-\alpha} - 1\right) - k_{2,\alpha} \right) \\
		\rho_2 &=  \gamma k_{1,\alpha} \mu^{-\alpha}
	\end{split}
\end{equation}
To compare $\dot{V}$ and $V$, Jensen's inequality, \cite{beckenbach2012inequalities}
is used and:
\begin{equation}
	\left(\dot{V}\right)^{\frac{\beta}{\alpha}} \geq \left( \left(\rho_2^{\frac{1}{\alpha}} \psi^2 \right)^\alpha + \left(\rho_1^\frac{1}{\alpha} \lvert s \rvert^\alpha\right)^{\frac{\beta}{\alpha}} \right)^\frac{\beta}{\alpha} \geq \rho_2^\frac{\beta}{\alpha} \psi^{2 \beta} + \rho_1^{\frac{\beta}{\alpha}} \lvert s \rvert^\beta 
\end{equation}
which is valid for any $\beta > \alpha$.
Consider the case $\frac{1}{2} \leq \alpha < 1$ and select $\beta = 1$. Then
\begin{align*}
		\left(-\dot{V}\right)^\frac{1}{\alpha} &\geq \rho_2^\frac{1}{\alpha} \psi^2 + \rho_1^\frac{1}{\alpha} \lvert s \rvert \geq \Gamma_1 \lambda_{max}(R) \psi^2 \nonumber\\
		&+ \Gamma_1( \lambda_{max}(R) \lvert s \rvert^{2 \alpha - 1} + \gamma k_{2,\alpha})\lvert s \rvert \geq \Gamma_1 V
\end{align*}
is valid in the $\mathcal{R}_1 = \{(s,\phi) \in \mathbb{R}^2 \mid \lvert s \rvert \geq S\}$ when:
\begin{align*}
	\Gamma_1 &\leq \min \left\{\frac{\rho_2^\frac{1}{\alpha}}{\lambda_{max}(R)}, \frac{\rho_1^\frac{1}{\alpha}}{\lambda_{max}(R) S^{2\alpha-1} + 2\gamma k_{2,\alpha}}\right\}
\end{align*}
This inequality can be satisfied for any $S \in \mathbb{R}^{*+}$. This implies that in any neighborhood of the origin the Lyapunov function satisfies the differential inequality $\dot{V} \leq -\Gamma_1^\alpha V^\alpha$, from which the finite time convergence follows \cite{bacciotti2005liapunov}.
Consider now the case $0 < \alpha \leq \frac{1}{2}$ and select $\beta = 2\alpha$. It follows that:
\begin{align*}
		\left(-\dot{V}\right)^2 &\geq \rho_2^2 \psi^{4\alpha} +  \rho_1^2 \lvert s \rvert^{2\alpha}\leq \Gamma_2 \lambda_{max}(R) \psi^2 \nonumber\\
		&+ \Gamma_2\left( \lambda_{max}(R) \lvert s \rvert^{2\alpha} + \gamma k_{2,\alpha} \lvert s \rvert \right) \nonumber\\
		&= \Gamma_2\left(\lambda_{max}(R) \psi^{2(1-2\alpha)}\right) \psi^{4\alpha} \nonumber\\
		&+ \Gamma_2 \left(\lambda_{max}(R + 2\gamma k_{2,\alpha} \lvert s \rvert^{1-2\alpha})\right)\lvert s \rvert^{2\alpha} \geq \Gamma_2 V
\end{align*}
is valid in the set $\mathcal{R}_2 = \left\{(s,\phi) \in \mathbb{R}^2 \mid \lvert s \rvert + \psi^2 \leq r^2\right\}$ when 
\begin{align*}
	\Gamma_2 &\leq \min \left\{\frac{\rho_1^2}{\lambda_{max}(R) \psi^{2(1-2\alpha)}}, \frac{\rho_2^2}{\lambda_{max}(R) + 2\gamma k_{2,\alpha} \lvert s \rvert^{1 - 2\alpha}}\right\}
\end{align*}
This inequality can be satisfied for any $r > 0$. In case $\alpha = \frac{1}{2}$ the inequality is valid globally. This implies that in any neighborhood of the origin (globally if $\alpha = \frac{1}{2}$) the Lyapunov function satisfies the differential inequality $\dot{V} \leq -\Gamma_2^\frac{1}{2} V^\frac{1}{2}$, from which the finite time convergence of the trajectories to the barrier set follows \cite{bacciotti2005liapunov}.

\noindent \underline{In the perturbed case ($\Delta \neq 0$)}, \eqref{eq:Lyap_Dynamics} can be rewritten as:
\begin{align}
	\dot{V} &\leq -\gamma \left(k_{1,\alpha} \left(k_{2,\alpha} - \left(1-\alpha\right)\mu^\frac{-\alpha}{1-\alpha} - 1 - \Delta \right) - k_{2,\alpha} \right)\lvert s \rvert^\alpha \nonumber\\
	&- \gamma k_{1,\alpha} \mu^{-\alpha}\lvert \psi \rvert^{2\alpha} +  2 \gamma \lvert \psi \rvert \Delta
	\label{eq:Lyap_dyn_pert}
\end{align}
Note that, $\lvert \psi \rvert$ is bounded on $\mathbb{R}^+$, indeed since $\lvert \psi \rvert = \lvert u + \delta(t) \rvert \leq \lvert u \rvert + \Delta$, as long as $\lvert u \rvert$ is bounded (which is always the case on real-systems), the derivative of the sliding variable is bounded. Hence, without loss of generality, it is possible to define $\Delta^{\prime} > 0$ such that $\Delta^{\prime} \geq \lvert \psi \rvert^{1 - 2\alpha} \Delta$ $\forall \alpha \in \left] 0 , 1 \right[$. Subsequently, \eqref{eq:Lyap_dyn_pert}, is rewritten as:
\begin{align*}
		\dot{V}& \leq -\gamma \left(k_{1,\alpha} \left(k_{2,\alpha} - \left(1-\alpha\right)\mu^\frac{-\alpha}{1-\alpha} - 1 - \Delta \right) - k_{2,\alpha} \right)\lvert s \rvert^\alpha \nonumber\\
		&- \gamma\left( k_{1,\alpha}\mu^{-\alpha} - 2\Delta^{\prime} \right)\lvert \psi \rvert^{2\alpha} \leq -\rho_1 \lvert s \rvert^\alpha -\rho_2 \lvert \psi \rvert^{2\alpha}
\end{align*}
with
\begin{align*}
		\rho_1 &= \gamma \left( k_{1,\alpha} \left(k_{2,\alpha} - \left(1-\alpha\right)\mu^\frac{-\alpha}{1-\alpha} - 1 - \Delta\right) - k_{2,\alpha} \right) \\
		\rho_2 &=  \gamma\left( k_{1,\alpha} \mu^{-\alpha} - 2\Delta^{\prime} \right)
\end{align*}
Considering that, $\gamma, \mu, \Delta \in \mathbb{R}^+$ and $\lvert s \rvert > \epsilon$, the gains are increasing as long as the following conditions aren't met:
\begin{align*}
		&\lvert s \rvert < \epsilon\\
		&k_{2,\alpha} > (1-\alpha) \mu^\frac{-\alpha}{1 - \alpha} - 1 -\Delta\\
		&k_{1,\alpha} > \max \left\{\frac{k_{2,\alpha}}{k_{2,\alpha} - (1-\alpha)\mu^\frac{-\alpha}{1-\alpha}-1-\Delta}, \frac{2\Delta^{\prime}}{\mu^{-\alpha}} \right\}
\end{align*}
Consequently, at a time $\tau$, both $\rho_1$ and $\rho_2$ are strictly positive. Therefore according to the second part of the unperturbed case stability proof, the perturbed system is also converging to the barrier set in finite time $\forall \alpha \in \left] 0, 1\right[$.
\end{proof}

\section{Multi-layered barrier function design} \label{sec:MultiLayer}
%
	In discrete-time systems, large perturbations acting on the system between two time samples can force $s$ out of the barrier set, denoted by $\vert s \vert \leq \epsilon_1$. When this happens, the system exclusively relies on the adaptive algorithm \eqref{eq:STA_dynamics} to maintain boundedness. This often leads to frequent and large variations of the controller gains and, eventually, over-actuation. To mitigate this issue, multiple barrier functions can be implemented, each one corresponding to different performance specifications.
    
	\subsection{Basic concept and mechanism}
		\begin{figure}[t] 
	\centering
	\includegraphics[width=0.45\textwidth]{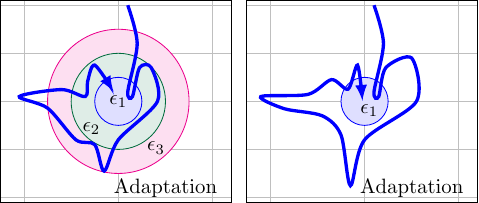}
	\caption{Conceptual illustration of three-layer barrier functions architecture (left) with step disturbances compared to single-layer barrier function (right). In the latter case the system solely relies on dynamic adaptation outside $\vert s \vert \leq \epsilon_1$.}
	\label{fig:SmultipleLayers}
\end{figure}

The key concept features a nested architecture of several layers, i.e. sets $\vert s \vert \leq \epsilon_i$ with corresponding barrier functions as illustrated in Fig. \ref{fig:SmultipleLayers}. The main components include:
\begin{itemize}
	\item The inner layer ($\epsilon_1$) corresponding to the primary barrier function intended to maintain $s$ within the desired accuracy tolerances.
	\item The outer layers ($\epsilon_i, \; i = 2,3,\dotsc, N$) corresponding to several, more relaxed barrier functions that act when $s$ is pushed out of the inner layer due to disturbances or large sampling intervals.
\end{itemize}
When a disturbance forces $s$ out of the inner set $\vert s \vert \leq \epsilon_1$, the trajectories enter the region governed by an outer barrier function corresponding to $\epsilon_i > \epsilon_1$. This outer barrier function constrains $s$ to this new, wider layer $\vert s \vert \leq \epsilon_i$ that corresponds to more relaxed performance specifications. While there, the  associated controller gains will ensure that $s$ enters the next inner layer if the perturbation bound does not change. The selection of the number $N > 0$ of outer layers and the associated bounds $\epsilon_i$ pertains to a reformulation of the control objective in each application with ``several levels of acceptable performance", where dynamic gain adaptation will only be employed when the performance has reached unacceptable levels ($\vert s \vert > \epsilon_i$). The selection of the bounds $\epsilon_i$ is based on the specifications associated to the levels of acceptable performance.
This layered approach can accommodate perturbations with varying bounds and slow sampling times compared to the size of the perturbation.

	\subsection{Multi-layered barrier function algorithm}
The proposed algorithm features two modes of modulation \( A_0 \) and \( A_i \) for the controller gains as shown in \eqref{eq:A0AiModulation}. These modes respectively correspond to dynamic adaptation ($k_{1,\alpha,d}$ and $k_{2,\alpha,d}$ are used) when $\vert s \vert > \epsilon_N$ and barrier function-based adaptation ($k_{1,\alpha}$ and $k_{2,\alpha}$ are used) when $\vert s \vert \leq \epsilon_i$.

\begin{align} \label{eq:A0AiModulation}
	&A_0 : \begin{cases}
		\dot{k}_{1,\alpha,d} = \displaystyle {k_{1,\alpha,d}}/{\lvert \dot{s} \rvert} \\
		\dot{k}_{2,\alpha,d} = \displaystyle \frac{k_{2,\alpha,d}}{2 \lvert s \rvert^{1-\alpha}} \\
		k_{1,\alpha} = k_{1,\alpha,d} \\
		k_{2,\alpha} = k_{2,\alpha,d}
	\end{cases}, A_i : \begin{cases}
		\dot{k}_{1,\alpha,d} = -k_{1,\alpha,d} \\
		\dot{k}_{2,\alpha,d} = -k_{2,\alpha,d} \\
		k_{1,\alpha} = \displaystyle \frac{\lvert s \rvert}{\left(\epsilon_i - \lvert s \rvert\right)^{\alpha+1}} \\
		k_{2,\alpha} = k_{1,\alpha}^2
	\end{cases}
\end{align} 
The conditions for switching between the two modes are based on the thresholds \( \epsilon_i \) with $\epsilon_N$ corresponding to the worst acceptable accuracy bound. The algorithm is formulated as:
\begin{align*} 
	\begin{cases}
		A_0 \text{ if } \lvert s \rvert > \epsilon_N \vee \left( \epsilon_N > \lvert s \rvert > \epsilon_1 \wedge  A_0 \text{ was in use} \right) \\
		A_i \text{ if } \epsilon_{i-1} < \lvert s \rvert < \epsilon_i  \wedge A_0 \text{ was not used } \forall i \in [\![2,N]\!]\\
		A_1 \text{ if } \lvert s \rvert < \epsilon_1
	\end{cases}
\end{align*} 
The condition inside the parenthesis corresponds to the case where the initial conditions are outside the largest allowable bound $\epsilon_N$. Note that the dynamic adaptation gains $k_{1,\alpha,d}$ and $k_{2,\alpha,d}$ decrease when the trajectories are within any barrier set, which prevents using large initial values for the gains from previous adaptations, reducing conservatism.

\section{Simulation} \label{sec:results}
	The performance of the proposed scheme, consisting of two layers ($N = 2$) with $\epsilon_1 = 10^{-4}$ and $\epsilon_N = 10^{-1}$, was assessed in simulation and compared to a single-layer adaptation solution. Both algorithms were implemented using Euler discretization. Two scenarios for system \eqref{eq:system} were considered with initial conditions outside the innermost barrier set.
\subsection{Sequence of steps}
\begin{figure}[t]
	\centering
	\includegraphics[width=0.45\textwidth]{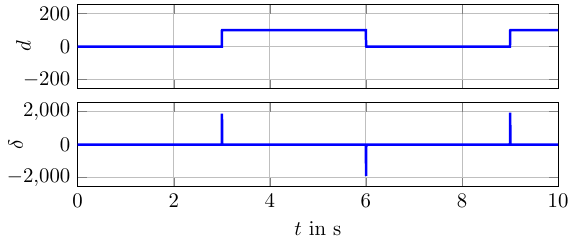}
    \vspace{-3pt}
	\caption{Perturbation as a sequence of step changes and its impulse-like derivative $\delta(t)$.}
	\label{fig:SimStepsDelta}
\end{figure}

In this scenario, the system is perturbed by square pulses of amplitude $d = 100$. This corresponds to high-amplitude (theoretically infinite) impulse profiles for $\delta(t)$ as shown in Fig. \ref{fig:SimStepsDelta}.
\begin{figure}[t]
	\centering
	\includegraphics[width=0.45\textwidth]{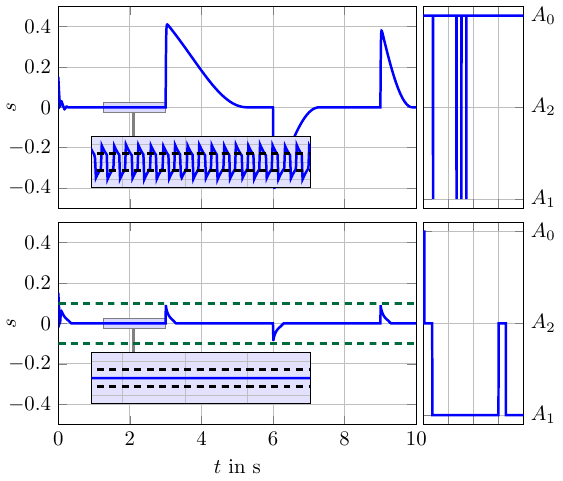}
    \vspace{-3pt}
	\caption{Sliding variable $s$ for step sequence perturbation with single barrier function (top) and double-layered architecture (bottom). The dashed black lines depict the inner layer $\epsilon_1$ and the dashed green lines depict the outer layer $\epsilon_2$.}
	\label{fig:SimSteps}
\end{figure}
\begin{figure}[t]
	\centering
	\includegraphics[width=0.45\textwidth]{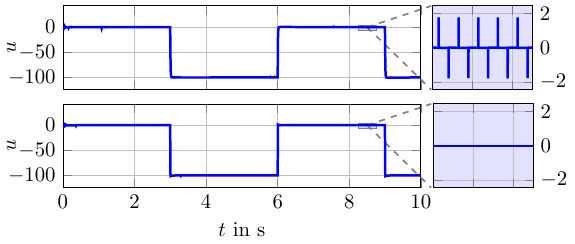}
    \vspace{-3pt}
	\caption{Control signal for sequence of step changes in single-barrier (top) and double-barrier (bottom) architecture.}
	\label{fig:SimStepsInp}
\end{figure}
The performance of the closed loop system is illustrated Fig. \ref{fig:SimSteps} for the single-layer (top) and double-layer (bottom) case. As it can be seen, the constant switching between dynamic and barrier-based gain adaptation in the single-layer architecture leads to excessive controller gains, as shown in the zoomed control signal plot in Fig. \ref{fig:SimStepsInp} (right hand-side subplots). This forces the sliding variable in and out of the inner barrier set. In the double-layer case, dynamic adaptation is only employed at the beginning ot the simulation since the system starts outside the outer barrier set. After one second, the trajectories are confined in the barrier set $\vert s \vert \leq \epsilon_2$ for all future time. In fact, apart some very small time interval after each step change, $s$ converges to the innermost barrier set, where it remains until the next perturbation step. This can also be seen from the right hand subplots in Fig. \ref{fig:SimSteps}, which illustrate the gain modulation modes.

\subsection{Varying frequency sinusoid}
\begin{figure}[t]
	\centering
	\includegraphics[width=0.45\textwidth]{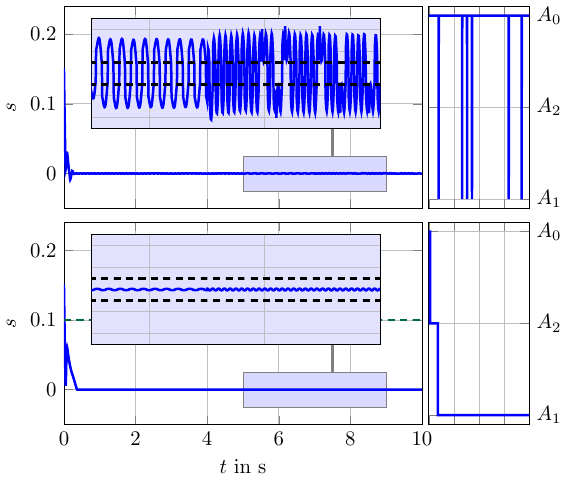}
    \vspace{-3pt}
	\caption{Sliding variable $s$ for sinusoidal perturbation with single barrier function (top) and double-layered architecture (bottom). The dashed black lines depict the inner layer $\epsilon_1$ and the dashed green lines depict the outer layer $\epsilon_2$.}
	\label{fig:SimSinus}
\end{figure}

\begin{figure}[t]
\centering
\includegraphics[width=0.45\textwidth]{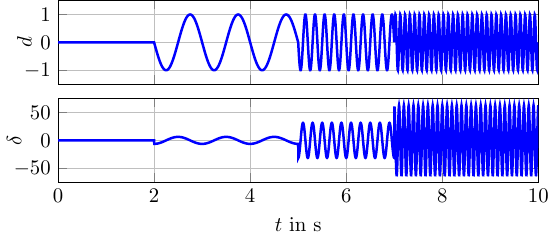}
\vspace{-3pt}
\caption{Sinusoidal perturbation $d(t)$ and its derivative $\delta(t)$ with varying frequency and amplitude.}
\label{fig:SimSinusDelta}
\end{figure}

This scenario introduces a sinusoid perturbation $d(t) = \sin(2\pi ft)$, where the frequency $f$ assumes the values 1Hz, 5Hz and 10Hz at times 2s, 5s and 7s, respectively. This implies that the perturbation derivative $\delta(t) = 2\pi f\cos(2\pi ft)$ is of varying frequency and amplitude, the latter assuming values close to 60 as illustrated in Fig. \ref{fig:SimSinusDelta}. In this scenario too, the double-layer scheme outperforms the single barrier function approach and maintains the sliding variable $s$ within the desired bounds, i.e. $\vert s \vert < \epsilon_1$ after some time as shown in Fig. \ref{fig:SimSinus}. Dynamic gain adaptation only takes place at the beginning of the simulation due to distant initial conditions. This is also verified from the right subplot showing the gain modulation regime of the algorithm. On the contrary, in the single-layer case, there is a frequent switching between the modes $A_0$ (dynamic adaptation) and $A_1$ (barrier-based modulation), which causes gain overestimation and, consequently, cannot keep the sliding variable inside the inner barrier set. At the same time it does not guarantee specific tolerances while operating out of the set.

\section{CONCLUSIONS AND FUTURE WORKS}
	A adaptive scheme of the generalised non-homogeneous STSMC was presented in this paper. The salient features of the proposed framework pertain to the use of positive semidefinite barrier functions for gain modulation and the employment of a multi-layer architecture that prevents over-estimation of the controller gains. The advantage of the proposed strategy is that it can ensure boundedness of the sliding variable associated to multiple performance specifications ranging from optimal to just acceptable. This feature is useful in cases when the perturbation rate bounds vary or are large compared to the sampling period and can be related to concepts of ``graceful degradation" in fault-tolerant systems. Future work will include extensive experimental evaluation of the presented framework and sensitivity analysis-based tuning of its meta-parameters, such as the number of layers. \label{sec:conclusions}
\vspace{-10pt}
\section*{ACKNOWLEDGMENTS}
This research was conducted at DTU Electro as part of the HyRel project, funded by the Danish Energy Technology Development and Demonstration Programm (EUDP), grant number 64022-1058. The authors appreciate their support. Leonid Fridman acknowledges the financial support of  the Programa de Apoyo a Proyectos de Investigacion e Innovacion Tecnologica, Direccion General de Asuntos del Personal Academico, Universidad Nacional Autonoma de Mexico under Project IT100625.


\balance
\bibliographystyle{IEEEtran}        
\bibliography{Bibliography/mybib} 

\end{document}